\documentclass[prb,aps,amssymb,showpacs,twocolumn,superscriptaddress]{revtex4-1}
\usepackage{amsmath}
\usepackage{amssymb}
\usepackage{amsthm}
\usepackage{amsfonts}
\usepackage{listings}
\usepackage{enumerate}
\usepackage{latexsym}
\usepackage[dvips]{graphicx}
\usepackage{psfrag}
\usepackage{bm}
\usepackage{graphicx}

\newcommand{\beq}{\begin{equation}}
\newcommand{\eneq}{\end{equation}}

\input{epsf}

\begin{document}

\tolerance 10000

\newcommand{\vk}{{\bf k}}

\title{From Irrational to Non-Unitary: on the Haffnian and Haldane-Rezayi wave functions}

\author{M. Hermanns}
\affiliation{Department of Physics, Princeton University, Princeton, NJ 08544, USA}
\author{N. Regnault} 
\affiliation{Laboratoire Pierre Aigrain, ENS and CNRS, 24 rue Lhomond, 75005 Paris, France}
\author{B. A. Bernevig}
\affiliation{Department of Physics, Princeton University, Princeton, NJ 08544, USA}
\author{E. Ardonne}
\affiliation{Nordita, Roslagstullsbacken 23, SE-106 91 Stockholm, Sweden}

\begin{abstract}
We study the Haffnian and Haldane-Rezayi quantum Hall wave functions and their quasihole excitations by means
of their `root configurations', and point out a close connection between these seemingly
different states. For both states, we formulate a `generalized Pauli-principle', which makes it possible
to count the degeneracies of these states. The connection between these states might
elucidate the underlying theory describing the `irrational' Haffnian state. 
\end{abstract}

\date{\today}


\maketitle

The characterization of topological phases of matter is
an inherently difficult problem, because non-local properties determine in what type of topological phase (if any) a system is. 
Indeed, having a better understanding of why certain models
fail to fully develop such a phase will improve our understanding of topological phases of
matter. In this paper, we study two particular model states of the fractional quantum Hall 
effect, the most celebrated experimental system exhibiting topological order. These two
model wave functions, the Haffnian and the Haldane-Rezayi (HR) states, do not, in fact,
 describe a topological phase of matter.
Although they have different properties at first glance, we show that these two states are in fact closely
related to one another. By studying this relationship, one comes
closer to answering the question of what precisely
constitutes a topological phase of matter.  

Although the two above mentioned wave functions do not describe topological states of matter,
we refer to them as `states' in the following.
The Haffnian, which is a d-wave paired state of spinless bosons
at filling fraction $\nu = 1/2$, was studied in detail in
[\onlinecite{thesis:green01}]. The wave function reads
$$
\Psi_{\rm Hf} = {\rm Hf} \bigl(1/(z_i-z_j)^2\bigr)
\prod_{i<j}(z_i-z_j)^2 \ ,$$
where ${\rm Hf}$ denotes the Haffnian of matrix. It is the unique densest zero energy ground state of 
the local three-body Hamiltonian that penalizes any triplet that has relative angular momentum
less than 4.
It has been argued\cite{thesis:green01} that the Haffnian describes a phase transition between
the incompressible bosonic Laughlin state and a gapped d-wave paired state. 

The HR state is a fermionic, spin-singlet d-wave paired state
$$
\Psi_{\rm HR} = \det \Bigl(1/(z^\uparrow_i-z^\downarrow_j)^2\Bigr)
\prod_{i<j}(z_i-z_j)^2 \ ,$$
where the $z^\uparrow_i$'s (resp. $z^\downarrow_i$'s) denote the position of particles with spin
up (resp. spin down), while the spin index is omitted 
when the product runs over all coordinates irrespective of the spin. The HR is the unique
densest zero energy ground state of a `hollow core' two-body Hamiltonian\cite{hr88}.
It was initially proposed as a spin-singlet candidate 
to explain the physics at $\nu=5/2$. Read and Green\cite{rg00}
argued, by means of a low wavelength mapping to a critical d-wave superconductor,
that the HR state describes the phase transition between a weak paring, d-wave
spin-singlet phase and a strong-pairing phase.

To explain why the Haffnian and HR states do not qualify as topological
phases, we consider the conformal field theory (CFT) descriptions of these
states. The CFT describing the HR state is (apart from the $U(1)$ part
describing the charge) a non-unitary CFT\cite{gfn97}. It has
been argued that wave functions described by non-unitary CFT's do not
describe topological phases (see, for instance, [\onlinecite{src07}]),
although a microscopic understanding of this failure is lacking. To describe
the problems with the Haffnian wave function (see also [\onlinecite{r09}]), we note
that the torus degeneracy, and hence the number of types of excitations,
grows with the number of particles, which is unphysical. 

In this paper, we focus on counting,
characterizing and finding a relation between the quasihole excitations of both
the Haffnian and the HR states through the means of 
the underlying generalized Pauli principles (or
Ôexclusion statisticsÕ\cite{h91}).
We find that previously defined Pauli principles\cite{pauliprinciple}
underestimate the counting of Haffnian quasihole, and extra exclusion
statistics rules must be imposed to obtain the correct counting. The
extra configurations present for the Haffnian, when generalized to the
spinfull case, also reproduce the correct counting of the HR quasiholes.

{\em Pauli principle: the Haffnian on the sphere -} 
We will start our investigations by examining the Haffnian state on the
sphere\cite{h83} pierced by $N_\phi$ flux quanta. In the lowest Landau level, there are 
$N_\phi+1$ single particle states that are eigenstates of angular momentum, with $l_z$ values 
ranging from $-N_\phi/2$ to $N_\phi/2$. Wave functions on the sphere are 
related to those on the plane using the stereographic projection.
The bosonic many-particle wave functions can be expanded on Fock basis
$\Psi = \sum_{\mu} c_\mu m_\mu$. The Fock states $m_\mu$  are labeled by
their occupation number configuration  $(n_{N_\phi/2},\ldots,n_{-N_\phi/2})$, where
$n_j$ is the occupation number of the single particle orbital with angular momentum $j$.
This formalism is valid for both bosons (where $m_\mu$ are monomials) and fermions
(where $m_\mu$ are slater determinants).

Many model quantum Hall states (all the ones that are zero modes of at least one
pseudopotential Hamiltonian)
have non-zero coefficients for only a part of the Hilbert space\cite{pauliprinciple},
beyond simple symmetry considerations such as their total
angular momentum projection $L_z$.
There exists a `root configuration' from which
all configurations of the Fock states with a priori non-zero coefficients can be obtained.
Obtaining these states is done by `squeezing', a procedure in which the relative
angular momentum of two particles is decreased by two,
relative to its value in the root partition, while the total angular momentum is kept
constant. By squeezing repeatedly from the `root configuration',
one obtains all the Fock states which can have non-zero coefficients in the model states
(the `reduced Hilbert space').
Interestingly, the root configuration for model quantum Hall states, corresponds to the
state surviving in the Tao-Thouless limit\cite{tao-thouless}.

To obtain the ground state $\Psi$, one imposes the condition that
$\Psi$ is an $L=0$ state;
because for ground states all basis states have $L_z=0$, it suffices to
require that $L^{+} \Psi = 0$. 
Ideally, one would like this procedure to give a {\em unique} solution for the
$c_\mu$ and hence a unique state, irrespective of the
number of particles. For many model states, including the Haffnian, this is
indeed the case. The Haffnian ground state for $N_{\rm e}$ particles, has
$N_\phi = 2 N_{\rm e} - 4$ flux quanta. The root configuration of the Haffnian state, 
$(2,0,0,0,2,0,\cdots,0,2,0,0,0,2)$, is the densest configuration
subject to the rule that there are maximally two particles in four orbitals.

Apart from the ground states, one can also obtain the states at higher flux
$N_\phi = 2 N_{\rm e} - 4 + n_{\rm qh}/2$, that is, in
the presence of $n_{\rm qh}$ quasi-holes, which show characteristic degeneracies.
The procedure mimics the procedure for the ground state.
One first constructs the `root configurations'
corresponding to the lowest weights of the possible $L$ multiplets.
For $N_e=6$, and one added flux quantum, the
root configurations needed are
\begin{align*}
(2,0,0,0,2,0,0,0,2,0)& & L_z &= 3\\
(2,0,0,0,2,0,0,0,1,1)&& L_z &= 2\\
(2,0,0,0,2,0,0,0,0,2)&& L_z &= 1\\
(2,0,0,0,1,1,0,0,0,2)&& L_z &= 0\\
\end{align*}
in which the particles are packed as dense as possible at high angular momentum.
The reduced Hilbert spaces are obtained by squeezing, and the number of states
at $L=L_z$ is given by the number of solutions for $c_\mu$ of $L^{+}\Psi=0$.
We checked that this procedure
is in full agreement with the counting of zero energy eigenstates of the model Hamiltonian,
as performed in [\onlinecite{thesis:green01}], resulting in the counting formula
\begin{equation}
\sum_{b}\binom{b-2+n_{qh}/2}{b}\binom{(N_e-b)/2+n_{qh}}{n_{qh}}\, .
\end{equation}

From the counting formula one can already see that the
Haffnian corresponds to an `irrational' theory, in which the number of excitations grows
with the number of electrons.
Summing the first factor $\binom{b-2+n_{qh}/2}{b}$ over $b$, one 
obtains part of the degeneracy on the torus\cite{footnote1}.
The sum over $b$ is only
constrained by the number of electrons, $b\leq N_e$ (and not by $n_{\rm qh}$ as is
the case for many other states), showing that the
degeneracy on the torus grows with the number of electrons, instead of being constant
for gapped quantum Hall states\cite{wn90}.

In order to obtain the number of  $L=L_z$ multiplets present at a chosen value of $N_\phi$,
we now introduce
a `generalized Pauli principle', which provides a way of
counting the number of states for $N_e$ particles, at a given flux $N_\phi$. Namely, one
writes all the `orbital occupation' configurations, subject to some rules.
In the case of the (bosonic) Read-Rezayi states, these rules are simply that no more than
$k$ particles can occupy two neighboring orbitals\cite{pauliprinciple}, and the number
of multiplets at a certain $L$ can be obtained as the difference between the number of
states at $L_z$, $L_z+1$ (see also [\onlinecite{rrrootconf}]).

For the Haffnian, the main rule is that no more than
two particles can occupy four consecutive orbitals. However, such a rule by itself is not
consistent with the counting, because it does not capture the `irrational' behavior
described above. We introduce an additional rule, which states that the pattern
`$0,2,0,0,1,0$' is also allowed. 
One way to view this additional rule is that one
allows in a $\nu= 1/2$ Laughlin-like root pattern `$1,0,1,0,1,0,1$', for the squeezing of two
`neighboring' particles, that is `$0,1,0,1,0$'$\rightarrow$`$0,0,2,0,0$', as long as one does
not generate a sequence `$0,1,0,0,2$'.
Alternatively, one can think of every configuration `$0,2,0,0,1$' as having to appear
symmetrized with the '$0,1,0,0,2$' configuration, thereby not counting the latter to
avoid double counting.
We have checked extensively that by counting the
configurations given by these rules, one does indeed generate the right number of states.
It is suggestive that the Laughlin patterns are the ones that give rise to patterns capturing
the irrational behavior - the Haffnian is at the same filling as the $1/2$ Laughlin state.

{\em  Dressing with spins: the Haldane-Rezayi case --}
We now move to the HR state, a \emph{fermionic}
d-wave singlet state, also at filling fraction $\nu = 1/2$. As we pointed out,
the HR state is also the unique zero energy ground state of
a model Hamiltonian, at flux $N_\phi = 2 N_{e} -4$. The procedure of defining this
state and its quasiholes, by generating the Hilbert space from squeezing of a
`root configuration', and subsequently demanding that one has an $L=0$ state that
can be generalized to the spin-full case. In this paper, we mainly state the
appropriate recipe\cite{footnote2},
which we motivate more thoroughly in a different publication\cite{spinsqueeze},
where we will deal with a plethora of spin-singlet states.

To define the HR state, both for the ground state at flux $N_\phi = 2N_{e}-4$,
as well as for quasi-hole states at flux $N_\phi = 2 N_{\rm e} - 4 + n_{\rm qh}/2$,
one follows this simple recipe:
start with the same root configurations used for the Haffnian state
(the HR root configuration was also  considered in [\onlinecite{mjv09}]).
From these root configurations, generate the
Hilbert space by squeezing, with the constraint that the occupation of each orbital is
maximally two, because we are dealing with spin-$1/2$ fermions.
For each configuration obtained, one adds spin to the particles, in all possible
ways consistent with the hollow-core Hamiltonian.
For the HR state, the relative angular momentum of two
electrons with the same spin is at least three.
General spin-full fermionic states can be written as
$\Psi = \sum_{\mu,\nu} c_{\mu,\nu} m_\mu (z^{\uparrow}_i) m_\nu (z^{\downarrow}_i)$,
in terms of the coordinates of the spin-up $z^{\uparrow}_i$ and down $z^{\downarrow}_i$
electrons.
To obtain all the $(L,S)$ multiplets, one generates the reduced Hilbert spaces for all
$S_z \geq 0$ and then imposes the highest weight angular momentum and total spin
constraints  $L^{+} \Psi = S^{+} \Psi  = 0$ to find all the multiplets.
We checked that this indeed yields all the states expected from the character formula,
which was obtained by studying the number of zero energy states of the model Hamiltonian
for arbitrary flux\cite{rr96}. 

For the HR state, we can also define a generalized Pauli principle: start from all the
configurations satisfying the
generalized Pauli principle for the Haffnian, and dress them with spin.
All orbitals occupied by two particles must harbor a singlet pair, because the particles
are fermions. The Hamiltonian implies that the
same is true for two nearest neighbor orbitals, as well as for two
next-nearest neighbor orbitals, which are both singly occupied.
Only when both neighbors and both next-nearest neighbors of an occupied orbital
are unoccupied, is the spin of the particle arbitrary, that is, the densest occupation around
a `free' spin is `$1,0,0,1,0,0,1$', where the middle particle corresponds to the `free' spin.
We note that as a result, configurations of the form `$1,0,1,0,1$' will be absent for the
HR state, as we would need to form singlets between $3$ spin $1/2$ particles.

The counting of states is now
simple: take all the Haffnian configurations dressed with spins in all possible ways consistent with
the spin-part of the Hamiltonian.
These configurations give rise to several spin-multiplets. In general, the
Hamiltonian forces several spins to be part of a
singlet based on their relative orbital distance.
The remaining spins are free, and the only task left is to determine how many
different $S$ multiplets (and their degeneracy) can be formed out of these free spin-1/2
particles. This last problem is standard, the number of spin-$s$ multiplets
present in the product of $n$ spin-1/2's is given by
$\frac{2s+1}{(n+2s)/2+1}\binom{n}{(n-2s)/2}$.
We have checked extensively that this Pauli-principle indeed gives rise to the
same number of $(L,S)$ multiplets as the analytical counting.

Apart from reproducing the correct state counting of the quasihole states, our 
generalized Pauli principle for the Haffnian and HR states also 
gives the correct prediction of the orbital entanglement level counting on the sphere
(introduced in [\onlinecite{lihaldane}], see [\onlinecite{hroes}] for the HR case),
as well as the particle entanglement spectrum\cite{particleentanglement}.

{\em The Torus geometry --}
The Pauli principles obtained above are valid on a genus $0$ geometry.
To further elucidate the connection between the Haffnian and HR states, we study
these states on the torus, and formulate the correct generalized Pauli principle.
We performed calculations using the translation symmetry along the $y$ direction.
Thus the states are eigenstate of the 
momentum along $y$ with values $K_y=\left(\sum_i n_i\right) {\rm mod} N_\phi$.

It is known that
the ground state degeneracy of the Haffnian, in the absence of quasi-holes, grows with
system size. In particular, the torus degeneracy is $N_e + 8$ or $N_e +1$ for $N_e$ even
or odd, as obtained by exact diagonalization\cite{rezayi}.
In contrast, for the HR state, the degeneracy is $10$ ($2$) for
$N_e$ even (odd), see for instance [\onlinecite{gfn97}].
We now see how this information comes out of the generalized Pauli principle.

The basic rule on which the generalized Pauli principle
is based - no more than two particles in four orbitals - gives rise,
for $N_e$ even, to ten states based on the configurations
\begin{align}
\nonumber
&(2,0,0,0,2,0,0,\ldots,2,0,0,0) & K_y &=  0,\frac{N_{e}}{2},0,\frac{N_{e}}{2}\\ 
\nonumber
&(1,1,0,0,1,1,0,\ldots,1,1,0,0) & K_y &= \frac{N_{e}}{4},\frac{3N_{e}}{4},\frac{N_{e}}{4},\frac{3N_{e}}{4} \\ 
&(1,0,1,0,1,0,1,\ldots,1,0,1,0) & K_y &= \frac{N_{e}}{2},0
\label{basicevenpatterns}
\end{align}
and their translations, occurring at the indicated momenta.
However, the torus lacks the "shift" of the sphere, and the Haffnian occurs at the same flux as the
$\nu=1/2$ Laughlin state, meaning that our second Pauli rule gives rise to multiple other configurations that are associated with the presence of the Laughlin partitions in Eq. (\ref{basicevenpatterns}).

To obtain all the torus ground states for
the Haffnian, we employ the same procedure as we did on the sphere, namely,
we allow configurations which contain patterns of the type `$0,2,0,0,1$',
provided they can be squeezed from the $\nu=1/2$ Laughlin pattern `$1,0,1,0,1$'.
Because of the periodic boundary conditions on the torus, we do need to allow one
occurrence of the pattern `$0,1,0,0,2$', if and only if no quasi-holes are present.
This one occurrence can be `located' in two different
positions. Thus, apart from the configurations we listed above, we have the
following additional configurations (for $N_e = 8$). Notice the presence of the
sequences `$1,0,0,2$' due to the periodic boundary conditions.
\begin{align}
(2,0,0,1,0,1,0,1,0,1,0,1,0,1,0,0) && K_y = 0 \nonumber\\
(2,0,0,0,2,0,0,1,0,1,0,1,0,1,0,0) && K_y = 0 \nonumber\\
(2,0,0,0,2,0,0,0,2,0,0,1,0,1,0,0) && K_y = 0 \nonumber\\
(0,2,0,0,1,0,1,0,1,0,1,0,1,0,1,0) && K_y = 8 \nonumber\\
(0,2,0,0,0,2,0,0,1,0,1,0,1,0,1,0) && K_y = 8 \nonumber\\
(0,2,0,0,0,2,0,0,0,2,0,0,1,0,1,0) && K_y = 8 \label{additionalevenpatterns}
\end{align}
In general, for $N_e$ even, we find $N_e-2$ additional states, for total of $N_e+8$
states.
For $N_e$ odd, we find a total of $N_e+1$ states, half of them are at $K_y=0$, the other
half at $K_y=N_\phi/2$.

Turning to the ground state degeneracy of the HR state, we follow the same
procedure as we did for the sphere, namely by dressing the Haffnian configurations with spin,
taking the constraints of the Hamiltonian into account. This means that configurations
which contain the pattern `$1,0,1,0,1$' are excluded, and for all even system sizes, we find
ten states, eight of which correspond to the first two lines of Eq. \eqref{basicevenpatterns},
the remaining two, which contain the pattern `$2,0,0,1$', correspond to lines three and six
of Eq. \eqref{additionalevenpatterns}. For $N_e$ odd, this procedure gives only two
ground states. All ground states have $S=0$.

We now briefly turn to the quasi-hole case, starting with the Haffnian.
Apart from allowing the configurations which
are characterized by allowing maximally two particles in four neighboring orbitals, we also allow
configurations which contain `$0,2,0,0,1$', as long as they can be obtained by squeezing
from `$1,0,1,0,1$', and do not contain the pattern `$0,1,0,0,2$'.
The configurations thus obtained are in one-to-one correspondence to the ground states of
the model Hamiltonian of the Haffnian state, which we checked by explicit diagonalization.
The configurations for the HR state are obtained from those of the Haffnian, by dressing them
with spin, in all ways consistent with the Hamiltonian.
This excludes patterns `$1,0,1,0,1$', and forces pairs of particles in
patterns of type `$0,0,2,0,0$', type `$0,0,1,1,0,0$' and type `$0,0,1,0,1,0,0$' to form
singlets. The remaining spins are free, and are allowed to form arbitrary spin multiplets,
whose counting we described above.
We confirmed the counting of torus states described here by explicit diagonalization
of the model Hamiltonian, and found complete agreement. As an example, we give
the results for $N_e=6$ particles, and $N_\phi = 12,\ldots,16$ (i.e. from zero until
four added flux quanta) in Table \ref{tab:N6counting} for both the HR and the Haffnian states.

\begin{table}[t]
\begin{tabular}{c || c | c c c c c c || c c c c c c}
&\multicolumn{7}{c}{Haldane-Rezayi} & \multicolumn{6}{c}{Haffnian} \\
$N_\phi$ & $S$ & $K_y=0$ & 1 & 2 & 3 & 4 & 5 & $K_y=0$ & 1 & 2 & 3 & 4 & 5 \\
\hline
12 & 0 & 3 & 0  & 0  & 2 & 0  & 0 & 5 & 0 & 0 & 2 & 0 & 0 \\
13 & 0 & 7 & - & - & - & - & - & 10 & - & - & - & - & - \\
14 & 0 & 28 & 26 & - & - & - & - & 40 & 37 & - & - & - & - \\
& 1 & 4 & 6 & - & - & - & - & & & & & & \\
15 & 0 & 75 & 72 & 72 & - & - & - & 102 & 99 & 99 & - & - & - \\
& 1 & 27 & 27 & 27 & - & - & - & & & & & & \\
16 & 0 & 165 & 160 & - & - & - & - & 214 & 208 & - & - & - & - \\
& 1 & 83 & 88 & - & - & - & - & & & & & & \\
& 2 & 2 & 2 & - & - & - & - & & & & & & \\
\end{tabular}
\caption{Number of multiplets for the HR and Haffnian states on the torus with
$N_e=6$ particles and $N_\phi = 12,\ldots,16$ flux quanta.
Dashes indicate repeated degeneracies enforced by symmetry, with momentum
$K_y$ period $\gcd (N_\phi,N_e)$.}
\label{tab:N6counting}
\end{table}

{\em Discussion --}
We uncovered a connection between the bosonic,
polarized d-wave paired Haffnian state, and the fermionic, spin-singlet d-wave
HR state, which both have filling fraction $\nu=1/2$. The connection
between these states is rather indirect, namely via their `root-configurations', which
encode important (topological) properties of these states. Although this connection
could well be `accidental', it might nevertheless shed light on the (irrational)
CFT underlying the Haffnian state. The CFT describing the
HR state is a non-unitary, $c=-2$ CFT\cite{hrcft}, which is closely related to a $c=1$ orbifold
theory\cite{gl98}. We leave the details of the CFT description for the Haffnian
(based on orbifold CFT's, see also [\onlinecite{bw10up}]) 
and its connection with the generalized Pauli principle for future work.

The Haffnian `root-configuration' is part of a general series,
namely $(2,0^{r-1},2,0^{r-1},2,\ldots,0^{r-1},2)$
with $r=4$. This series contains the Moore-Read state at
$r=2$,\cite{mr91} and the `Gaffnian'\cite{src07} at $r=3$. The latter correspond to a non-unitary CFT,
but one can re-interpret the root-configuration as one for spin-full fermions,
and construct a different state. In this case, one finds the `spin-charge separated' 
state\cite{all02}, whose non-abelian statistics is of Ising type. Interestingly,
this state is described by a unitary CFT, in contrast to the `Gaffnian'. As was the case
for the Haffnian, by increasing the internal degree of freedom, it is possible
to cure some of the problems plaguing the parent state. 

\emph{Acknowledgements}
We thank E. Rezayi for providing useful information on the torus geometry.
We thank the Aspen Center for Physics for its hospitality. NR was supported by 
the Agence Nationale de la Recherche under Grant No. ANR-JCJC-0003-01 and MH was supported by the Alexander-von-Humboldt foundation and NSF DMR grant 0952428. 
BAB was supported by Princeton Startup Funds, Alfred P. Sloan Foundation, NSF CAREER DMR- 095242, and NSF China 11050110420, and MRSEC grant at Princeton University, NSF DMR-0819860. BAB and MH wish to
thank Station Q for generous hosting during the last stages of preparation of this work.


\begin{thebibliography}{100}
\bibitem{footnote2}
We note that in principle, there are several ways of generalizing the squeezing
procedure to the spin-full case. Here, we presented a procedure which works for
all two-component states, and do not assume $SU(2)$ invariance.
\end{thebibliography}
\end{document}